\documentclass[twocolumn,10pt]{asme2ej}

\usepackage{epsfig} 
\usepackage{mathrsfs} 
\usepackage{subcaption}
\title{Growth of a Bubble Under a Fluctuating Pressure Field}

\author{V. Jukanti \& V. Srinivasan
    \affiliation{
        Multiphase Transport Phenomena Laboratory\\
	Department of Mechanical Engineering\\
	University of Minnesota\\
	Minneapolis MN 55455\\
    }
}

\begin{document}

\maketitle

\begin{abstract}
{\it This study seeks to understand the origins of intermittency in quantities of interest in pool boiling, such as bubble departure diameter and departure time. The intermittency of nucleation site activity due to nonuniform and unsteady near-wall temperatures is well-established; however few mechanistic models have been developed that predict such intermittency. Here we assume a fluctuating pressure field at a nucleation site due to adjacent bubble activity, which acts in combination with convective effects to alter bubble growth depending on the phase of the pressure oscillation with respect to the instant of bubble nucleation. The results suggest that even when a single departure frequency is assumed for nearby bubbles, its effect on the nucleation site being considered is to cause aperiodicity and intermittency in bubble departure quantities. The effects of pressure field phase angle, degree of superheat, and choice of force balance model on the bubble departure quantities are examined. The phase angle of the pressure fluctuation at the instant of bubble nucleation is shown to play a major role in determining bubble departure diameter and growth time. Departure diameter is observed to have a broad distribution over long times of observation, belying the assumption of a unique value. Period doubling of the ebullition cycle is observed for some conditions, a phenomenon documented by other investigators. The effects of dynamic contact angle  
}

\end{abstract}

\begin{nomenclature}
\entry{A}{Amplitude of sinusoidal wave}
\entry{$C_d$}{Drag force constant for H.Beer model}
\entry{$C_s$}{Drag force constant for Zeng and Klausner model}
\entry{$f$}{Frequency of the bubble departure}
\entry{$F_b$}{Buoyancy force}
\entry{$F_d$}{Drag force}
\entry{$F_s$}{Surface Tension force}
\entry{$F_p$}{Pressure difference force}
\entry{$F_i$}{Fluid inertial force}
\entry{$F_{net}$}{net force on the bubble}
\entry{$g$}{Acceleration due to gravity}
\entry{$h_{lv}$}{Latent heat vaporization}
\entry{$Ja_{T}$}{Jakob number based on initial superheat}
\entry{$K_{s}$}{Sphericity Correction.}
\entry{k}{Thermal conductivity of the liquid}
\entry{$m_{b}$}{mass of the vapor bubble}
\entry{R}{Instantaneous bubble radius}
\entry{$\Delta p_v$}{Additional dynamic vapor pressure drop}
\entry{$\Delta T$}{Initial superheat}
\entry{$T$}{Temperature of the bubble interface}
\entry{$t_w$}{Waiting time}
\entry{t}{Time}
\entry{\textbf{Symbols}}{}
\entry{$\alpha_{l}$}{Thermal diffusivity of the liquid.}
\entry{$\theta$}{Static contact angle.}
\entry{$\theta_D$}{Dynamic contact angle.}
\entry{$\nu_l$}{Dynamic Viscosity of liquid.}
\entry{$\rho$}{Fluid density}
\entry{$\mathscr{R}$}{Gas constant.}
\entry{$\sigma$}{Surface Tension.}
\entry{\textbf{Subscripts}}{}
\entry{l}{Liquid}
\entry{0}{Initial value of quantity}
\entry{v}{Vapor}

\end{nomenclature}

\section{Introduction}

The nucleate boiling regime of pool boiling heat transfer is of immense interest to thermal engineers because of the potential for high heat dissipation at high heat transfer coefficients, as well as the relatively simple geometric configuration that lends itself to engineering solutions such as immersion cooling. At the same time, it is also of fundamental interest to researchers interested in predicting how the application of constant heat flux or temperature boundary conditions to a surface leads to rich phenomena such as bubble nucleation, growth and departure from a subset of potential nucleation sites, along with critical phenomena such as sharp transitions to other modes of boiling such as film or transition boiling. \\

The determination of heat transfer for a certain surface heat flux (or temperature) is of primary interest to the engineer, and this requires knowledge of the active nucleation site density, the bubble departure diameter and the bubble departure frequency.  Many mechanistic models have been developed to predict these quantities  \cite{Han1965a,Han1965b,Ivey1967,Benjamin1996,zhang2021}; an excellent summary of the correlations for these quantities is given in the review of Pioro and Rohsenow \cite{Pioro2004a}. Such models usually calculate bubble departure diameter and frequency for a single bubble and differ in terms of the forces that are assumed to be acting on the bubble as it grows, along with some empirical constants. These models have provided considerable insights into the dynamics of bubble growth. It is now well understood that after a bubble departure event, there is a waiting time during which wall-adjacent fluid gets superheated to a sufficient extent that a cavity with trapped vapor can serve as a nucleation site. Immediately after nucleation, the small size of the bubble and associated high internal pressure leads to a rapid expansion, during which surrounding liquid is pushed back, likely causing convective motion around the bubble. After the bubble has achieved a size such that internal pressure effects are not dominant, the further growth of the bubble is controlled by heat transfer associated with the evaporation of the superheated liquid all around the bubble's surface area into the bubble volume (macro-layer evaporation) as well as near-wall superheated liquid (micro-layer evaporation). Bubble departure ensues when the upward forces, dominated by buoyancy, overcome other forces resisting motion. Understanding of the dynamics of a single bubble is then transferred to the entire population of nucleation sites, assuming that all cavities encounter similar conditions (Kolev\cite{Kolev1994}).

Many experimental measurements refute this assumption. Kenning and co-workers  \cite{kenning1996, Kenning2006} used liquid crystal thermography to measure temperature distributions on the underside of a thin heater foil with pool boiling above, and showed strong temporal and spatial variations of temperature. The temperature at a nucleation site depended on the history of temperature at that site, as well as those of nearby nucleation sites. This results in a dynamic distribution of nucleation sites, marked by intermittency of nucleation at individual sites. These findings were strongly in line with the computations of Sadasivan et al. \cite{Sadasivan1995}  who found intermittency of nucleation, and that adjacent cavities were activated when ebullition stopped at a nucleation site. Those authors also noted that at low heat fluxes, bubble departure was periodic, but the period doubled at higher heat fluxes, while departures became aperiodic at still higher heat fluxes, which is a classical marker of chaotic dynamics. Theofanous et al. \cite{Theofanous2002, Theofanous2002a} observed such non-periodic behavior through X-Ray and Infrared measurements of `reversible' and `irreversible' hotspots. Zhang and Shoji \cite{Zhang2003} documented the effects of cavity spacing on nucleation and attempted to assess the relative contributions of solid-solid, solid-liquid, and liquid-liquid interactions between nucleation sites. The chaotic dynamical aspect of boiling has also been implicitly incorporated into percolative models for the spreading of dry spots near Critical Heat Flux by Bucci \cite{Zhang2019}. \\

It should be noted that in addition to assuming spatial uniformity, which allows the extension of results for a single bubble to the entire surface, most mechanistic models also assume that conditions are steady during the growth process. However, from the discussion above, it can be seen that bubble growth is likely to occur in a highly unsteady field, due to the effects of convective motions from adjacent bubble departures, as well as changes in dynamic pressure due to the bubble departures in the surrounding liquid. Mechanistic models of bubble growth in unsteady fields are sparse in the literature. Jones and Zuber \cite{Jones1978} considered the growth of a spherical bubble in an infinite pool when the system experienced a sharp drop in pressure, such as in a steam nozzle. Wang and Bankoff \cite{Wang1991} performed similar calculations when a hemispherical bubble was located on an unheated surface. It was shown that bubble growth proceeded rapidly and departure diameters were larger under such conditions, as compared to the case of constant pressure. \\

Despite these advances, incorporation of the idea of bubble intermittency into mechanistic models remains elusive. Full-scale Direct Numerical Simulations \cite{Mukherjee2004, Dhir2006} have been shown to accurately reproduce visual observations of bubble growth and merger, but these are currently limited to domains containing a few nucleation sites. While these will undoubtedly improve in fidelity and span a larger range of spatial and temporal scales, they cannot currently replace the need for simpler models that reproduce observed behavior. Recently, the aperiodicity of heat flux fluctuations in pool boiling experiments aboard the International Space Station \cite{Raj2012, Raj2009} and corresponding terrestrial experiments [refs] were analyzed using a nonlinear dynamics framework. It was shown that heat flux fluctuations were non-Gaussian and non-stationary, exhibiting long-term temporal correlations. These temporal correlations were used to construct a universal curve for the Hurst exponent as a function of non-dimensional wall superheat that appeared to collapse data at multiple levels of system pressure, subcooling, and to some extent, gravity level \cite{Saini2021}. The present study seeks to build on that work by inquiring into the origins of the aperiodicity and seeks to develop a simple model that reproduces the aperiodic behavior of bubble departure quantities such as bubble departure diameter and departure time. \\

The goal is not to be predictive in the sense of being an engineering tool but to improve understanding. The basic premise is that bubble growth processes at one nucleation site are influenced by processes at nearby nucleation sites. This can happen through cooling of the substrate (thermal interactions), and convective motions set up by adjacent bubble departures (liquid-liquid and solid-liquid interactions). We model a subset of these interactions in this study. If bubble departures at all nearby nucleation sites were perfectly periodic, bubble growth at a site under consideration will occur in a fluctuating pressure field that is also marked by convective motion that can disrupt the superheated liquid layer around the bubble. We study the influence of these effects on the statistics of bubble growth time and departure diameter. 

\section{Problem Formulation}

\begin{figure}[h]\label{fig:schematic}
\centering
\includegraphics[width=\linewidth]{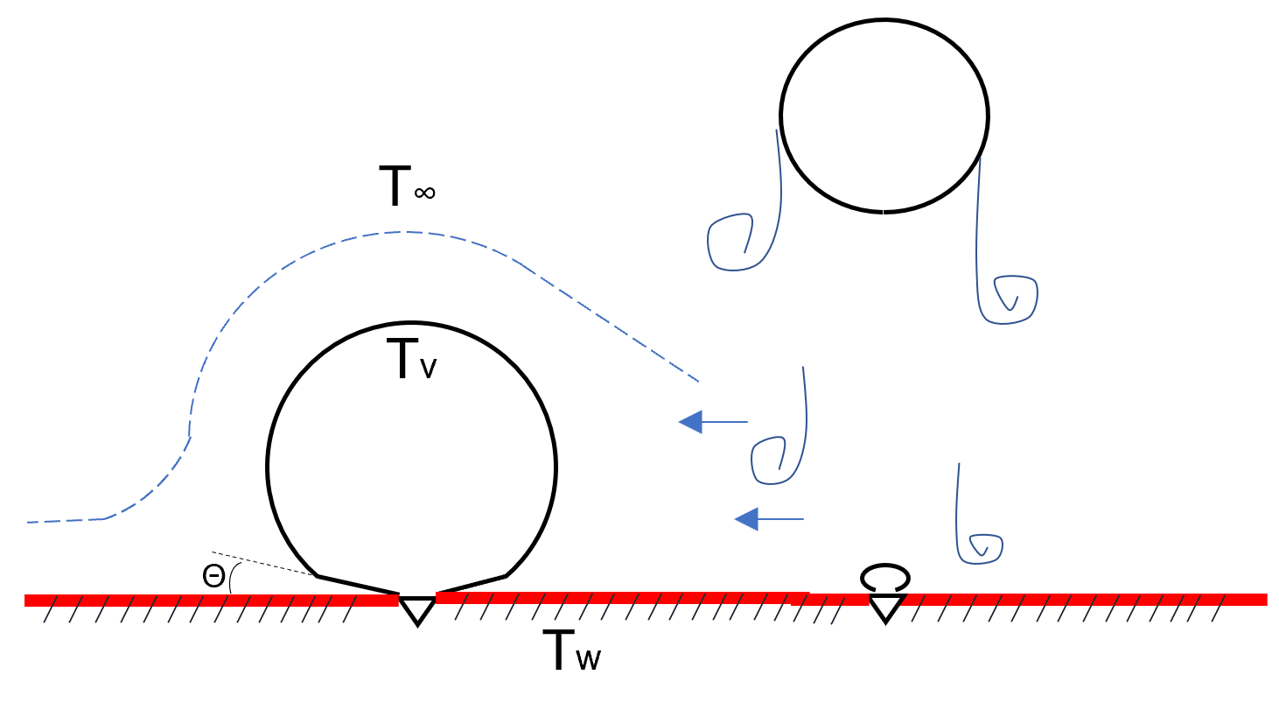}
\caption{Sketch of a bubble growing under an unsteady pressure and temperature field, with contact angle $\theta$ and a superheated macrolayaer around it.}
\label{fig:sketch}
\end{figure}

Before formulating the problem in the present study, it is useful to review the assumptions in the models of \cite{Jones1978} and \cite{Wang1991}, which the current model extends. For a time-dependent system pressure, Jones and Zuber \cite{Jones1978} made the key assumptions that allow for an analytical solution: (i) the bubble is in the thermally-controlled regime, since the inertial growth period occupies a very small fraction of the overall growth process (ii) this further implies that bubble radii are large enough that surface tension effects on bubble pressure can be neglected, (iii) the accommodation coefficient at the liquid-vapor interface is unity, and the bubble internal pressure is the same as the external imposed pressure, since sound speeds are high relative to fluid velocities, (iv) the vapor in the bubble is always saturated conditions, with the liquid-vapor interface temperature equal to that of the instantaneous saturation temperature for the instantaneous pressure. 

The rate of growth of the bubble can be written as 
\begin{equation}   
\frac{dm_{b}}{dt} = \frac{4\pi R^2\phi}{h_{lv}}
\end{equation}
where $\phi$ is the heat flux at the liquid-vapor interface driven by conduction between the far-field temperature and the interface temperature and needs to be specified to solve for R(t). This equation can be solved if bubble pressure and density can be specified as a function of time. To do this, they make the further assumption, valid at low reduced pressures, that the vapor density can be found from the ideal gas law and the Clausius-Clapeyron equation. 

With these assumptions, it can be shown that the problem reduces to an analysis of conduction in a semi-infinite medium (the liquid) with a time-dependent temperature imposed at one bounding surface (the phase interface). Further, the solution in spherical coordinates can be obtained by solving the problem in Cartesian coordinates and then applying a sphericity factor of $\sqrt{3}$. Starting from an initial pressure and temperature, a sudden reduction in system pressure is assumed to cause interface temperature variation of the form 
\begin{equation}\label{eq:temp variation}
    T_i(t)= T_0+F(t)   
\end{equation}

This results in the bubble being continuously superheated, leading to faster expansion than for constant conditions.  The result is 
\begin{eqnarray}{\nonumber}\label{eq:main}
\frac{R(t)}{\left( \frac{\rho _{v0}}{\rho _v}\right)^{1/3}} & = R_0 + \frac{K_sk_l\Delta T_s}{\rho _{v0}h_{lv}\sqrt{\pi \alpha _l}} \int _0^t   \left( \frac{\rho _{v0}}{\rho _v}\right)^{2/3} \eta ^{-1/2}d\eta \\ 
& -\frac{K_sk_l}{\rho _{v0}h_{lv}\sqrt{\pi \alpha _l}}\int _0^t   \left( \frac{\rho _{v0}}{\rho _v}\right)^{2/3}\int _0^\eta
\frac{F'({\tau})}{\sqrt{\eta -\tau}}d\tau d\eta
\end{eqnarray}

The first integral represents the constant-condition thermal growth solution while the second integral represents the contribution due to fluctuating pressure. Jones and Zuber \cite{Jones1978} show that if the saturation temperature decreases as $\sim t^n$ due to depressurization, then the bubble radius grows faster than for constant superheat as $R(t)\sim t^{n+1/2}$. Wang and Bankoff \cite{Wang1991} extended this idea to a bubble with contact angle $\theta$ on a spherical unheated wall and added the effects of microlayer evaporation underneath the bubble. They retained the sphericity factor of $\sqrt{3}$ for the curved surface evaporative heat transfer. The result is similar to Eq(\ref{eq:main}), except that the prefactor multiplying the integrals becomes a function of the contact angle $\theta$. For further details of the analysis, the reader is referred to refs \cite{Wang1991}. 

In the present model, we modify the above analyses to account for the fluctuating pressure field around a bubble under nominally constant conditions. It is first noted that during the inertial growth of a bubble, the interface velocity resulting from the solution of the Rayleigh-Plesset equation is given by 
\begin{equation}\label{eq:vel}
\dot{R} = \sqrt{\frac{2h_{lv}\rho _v}{3\rho _l}\left(\frac{T_\infty -T_{sat}(P_\infty)}{T_{sat}(P_\infty)}\right)}
\end{equation}

This interface velocity can be quite high ($\sim 5 m.s$ for water at 1 atm and 10K superheat), even though it may persist for only microseconds and the bubble diameter may be a few microns. This velocity will be converted into a pressure rise as the fluid moves away from the interface, leading to a periodic pulse felt by nearby bubbles. Further, it is well known that the bubble departure process creates a wake region with low pressure in the aft region of the bubble. Nearby bubbles will experience fluctuating pressure fields due to such departures. Kolev    \cite{Kolev1994} has estimated a scale for such velocities in the region between two bubbles. These effects are sought to be modeled by assuming that the pressure inside the bubble responds to such external fluctuations while retaining the assumption of saturation conditions. We thus have
\begin{equation}\label{eq:pressure variation}
P_v = P_l + \Delta Psin(\omega t+ \psi)
\end{equation}
or equivalently
\begin{equation}\label{eq:int temp var}
    T_v=T_0 +F(t) = T_0 - Asin(\omega t+ \psi)
\end{equation}
where choices need to be made for the amplitude A and the frequency of adjacent bubble departure $\omega$
These expressions are substituted into the integral equation for computing the bubble radius under thermally-controlled growth, Eq(\ref{eq:main}) to get the bubble radius R(t). A MATLAB code was written to evaluate the above expressions with second-order accuracy in time. Time steps in the range 5-100 $\mu s$ were used for the integration process, and it was verified that the behavior of R(t) was insensitive to time steps smaller than 10$\mu s$, which was then used for all calculations. 
\section{Results}
\subsection{Code Validation}
\begin{figure}
\centering
\includegraphics[width=\linewidth]{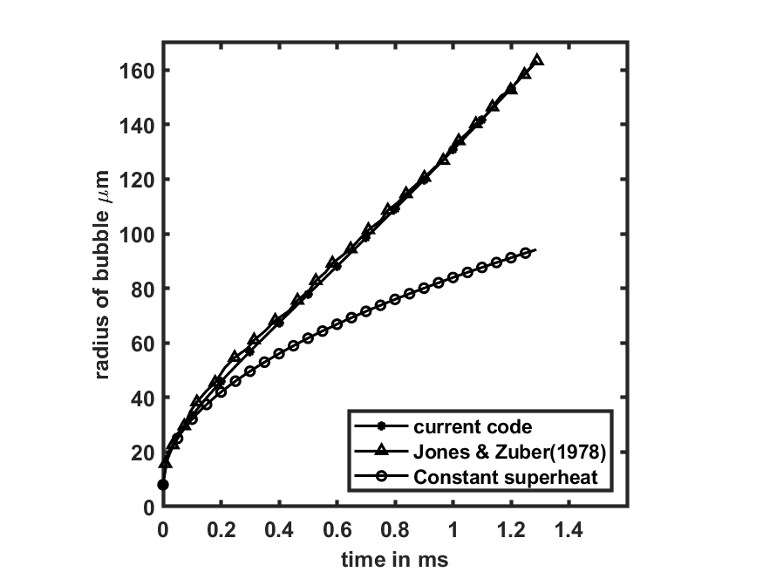}
\caption{Comparison of code results for growth of a spherical bubble with those of Jones and Zuber \cite{Jones1978} for a rapidly depressurizing pool with a linear decrease in saturation temperature. The constant pressure case is shown for reference.}
\label{fig:comp_JZ}
\end{figure}

The code is first validated for a transient pressure condition by replicating the results of Jones and Zuber \cite{Jones1978} for the case of pressure decreasing linearly with time (Fig. \ref{fig:comp_JZ}). For the case of constant pressure, the code matches the $\sqrt{t}$ behavior expected of thermally controlled growth \cite{Forster1954}. For a linear decrease, $F(t) =-bt$ in Eq(\ref{eq:main}),  the results match those in \cite{Jones1978} and show a $R\sim t^{3/2}$ behavior. Here, the initial condition for the bubble corresponds to the late stages of inertial growth. It is assumed that the bubble is about 10 microns in diameter after an extremely rapid phase of growth, and $t=0$ corresponds to conditions where thermally controlled growth can be assumed. 

\begin{figure}
\centering
\includegraphics[width=\linewidth]{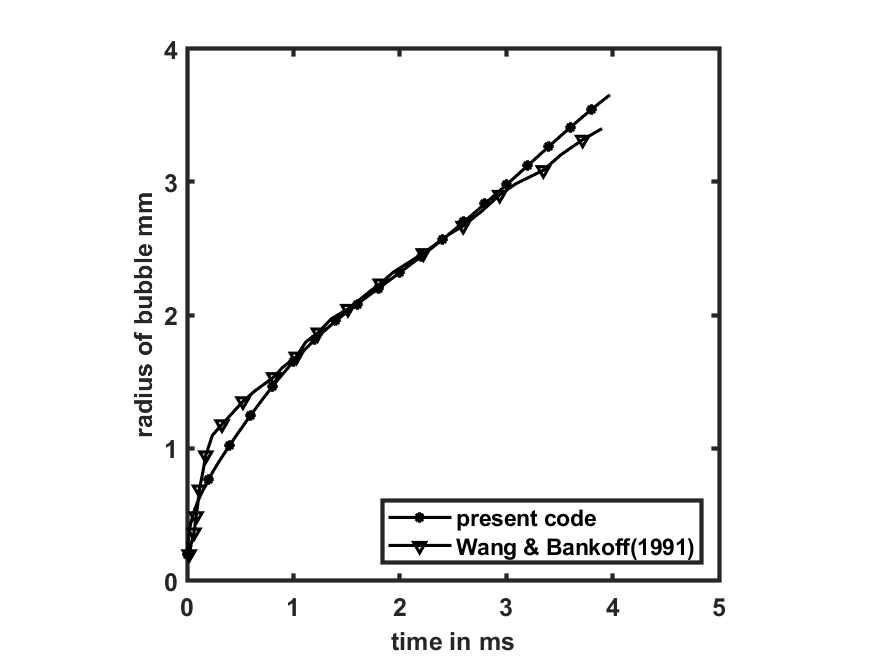}
\caption{Comparison of code results for growth of a hemispherical bubble cap on a wall with  those of Wang and Bankoff \cite{Wang1991} for a rapidly depressurizing pool}
\label{fig:comp_WB}
\end{figure}

Turning attention to the case of a bubble growing on a wall with a contact angle of $\theta$. Here the bubble volume is given by 
\begin{equation}
V_b= \frac{\pi}{3}R^3 (t)\left[ 2+cos\theta ( 2+sin^2 \theta )\right]
\end{equation}
and two different heat fluxes are defined, one denoted $\phi _c$ for the curved surface area and another $\phi _b$ for the microlayer underneath the bubble base. The quantities $\phi _c$ and $\phi _b$ are evaluated in the same way as $\phi$, however, $\phi _c$ is multiplied by the sphericity factor $\sqrt{3}$ to convert it for the spherical geometry from Cartesian coordinates. Figure ~\ref{fig:comp_WB} shows the comparison between the present results and those of \cite{Wang1991} for contact angle, $\theta =70^0 $ and a ninth-order polynomial fit to the pressure variation given by those authors. The agreement is within 4\% at the later stages and differences in early stages may possibly be due to experimental uncertainty in pressure measurement and fitting constants in their ninth-order fit. 

\subsection{Forces on Bubbles in Unsteady Pressure Fields}
In order to use these bubble growth models to determine departure diameters, we consider two force models that are widely used in the literature. While Kolev \cite{Kolev1994} has argued that under thermally-controlled growth corresponding to $R(t) \sim t^{1/2}$, the force balance leads to a static force equilibrium and dynamical forces dependent on the bubble or fluid acceleration need not be considered, this may change when the growth has a different time-dependence. Accordingly, the two models chosen differ in the number of force terms. The model of Klausner and co-workers\cite{Zeng1993}   considers only buoyancy acting upwards, and drag and surface tension forces acting downwards. The model of Beer and Best \cite{Beer1977} also includes forces corresponding to bubble inertia due to the motion of the center of gravity as well as the net pressure difference. For constant pressure conditions,  conditions of water at 1  atm. pressure and degree of super heat of 10K, the two models(\cite{Zeng1993},\cite{Beer1977}) give similar results for departure diameters of 3.40 mm and 3.35 mm respectively. This is primarily because the inertia term indeed does not appear to be significant, as argued by \cite{Kolev1994}. These models will be evaluated in the subsequent section for variable pressure fields. Further, a growth time to departure can be estimated, which can be combined with the waiting time of liquid superheat to arrive at a total time for the bubble life cycle, which can be inverted to form the bubble departure frequency. The values of bubble departure frequency from these two models for water at 1 atm are $~40$ Hz and $~42$ Hz respectively. 

\begin{figure}
\centering
\includegraphics[width=\linewidth]{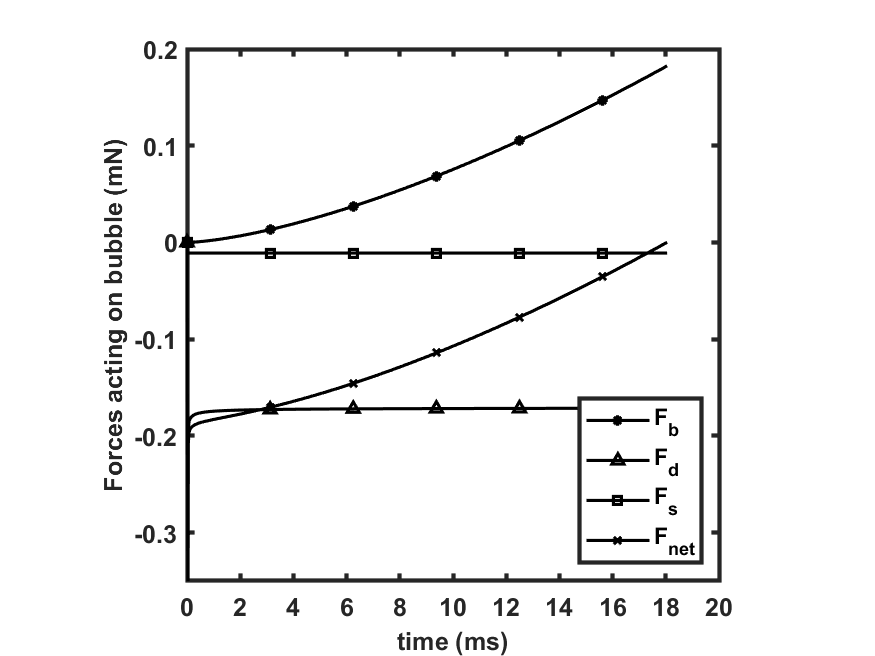}
\caption{Forces acting on the bubble during growth under a constant pressure field, using the model of Zeng et al. \cite{Zeng1993}.}
\label{fig:forces_klausner_const}
\end{figure}

\begin{figure}
\centering
\includegraphics[width=\linewidth]{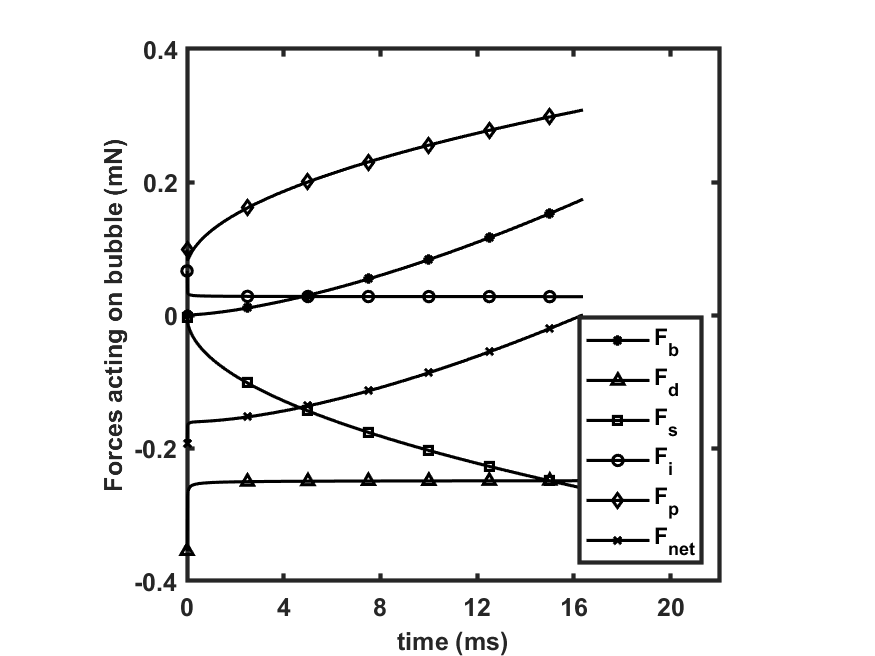}
\caption{Forces acting on the bubble during growth under an constant pressure field, using the model of Beer and Best \cite{Beer1977} }
\label{fig:forces_Beer_const}
\end{figure}

\subsection{Effects of an oscillating pressure field}

\begin{figure}
\centering
\includegraphics[width=\linewidth]{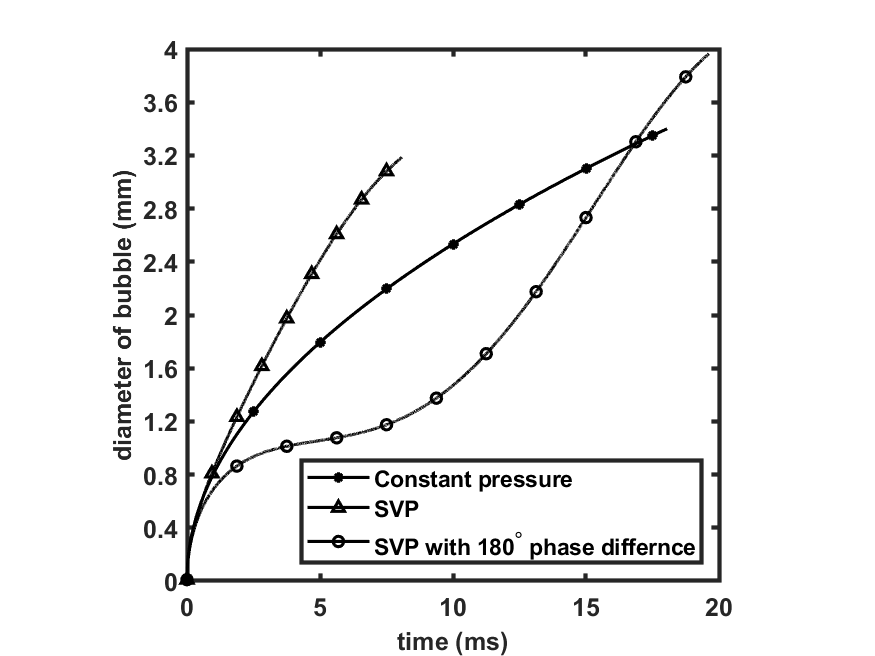}
\caption{Bubble growth for a sinusoidal oscillation of pressure (saturation temperature) with phase angles of 0$^o$ and $180^o$, compared to the constant condition situation.}
\label{fig:growth_vs_phase}
\end{figure}

The effects of an oscillating pressure field around a bubble are now explored for water at 1 atm, a superheat of 10K and a pressure oscillation frequency of 40 Hz. The pressure inside the bubble is assumed to vary sinusoidally, and together with the assumption of saturated vapor conditions, leads to a sinusoidal temperature variation given by Eq(\ref{eq:int temp var}). Fig. \ref{fig:growth_vs_phase} shows the effects of the phase angle of the oscillation on bubble growth. When the phase angle is zero, the pressure(saturation temperature) is dropping, and the bubble grows faster relative to the constant pressure situation. On the other hand, when the phase angle is $180^o$, the pressure (saturation temperature) is rising, which hinders the growth of the bubble. After  5.5 ms of time has elapsed and the phase angle is close to zero, the bubble growth becomes more rapid. The net result is that the bubble departure time is altered by the phase of the pressure oscillation frequency. Implicitly, this means that the presence of neighboring bubbles, and the precise time history of nucleation at these sites may alter the departure time of observed bubbles. 

When forces acting on the bubble are evaluated, the resulting picture is substantially different from the constant pressure situation. The behavior of the drag force is significantly altered, going from a nearly time-independent behavior for constant conditions to one in which it reaches a maximum in magnitude before decreasing. Comparison of Figs. \ref{fig:forces_Beer_const} and \ref{fig:forces_Beer_sin} show that the model of Beer and Best \cite{Beer1977} predicts a strong increase in the bubble inertia force and change in behavior of drag force under time-dependent pressure when compared to the case of constant pressure. The model of Zeng et al. \cite{Zeng1993} does not have the inertia term but its effects are captured by the drag force itself; similarly, it also captures a change in the behavior of the drag force when compared with the constant pressure case. The effect of the initial phase angle and the delay in bubble departure for a phase angle of $180^o$ is also illustrated using the force model of Zeng et al. \cite{Zeng1993} in Fig. \ref{fig:forces_Klausner_sin180}. It is seen that all forces are relatively weak because of slow initial bubble growth, and gradually increase in magnitude as the phase angle increases by $180^o$ back to $360^o$.  While both models predict a sharp drop in the time to bubble departure for a phase angle of $0^o$, they differ in predicted values: model \cite{Zeng1993} predicts a time of 8.08s while the model of \cite{Beer1977} predicts 6.50s. Despite the above differences in the forces, the departure diameters predicted by both models are very close, as seen in the following tables. 

\begin{table}[ht]
\caption{Constant pressure case } 
\centering 
\begin{tabular}{c c c } 
\hline\hline 
Force model & Departure diameter & Departure time \\ [0.5ex] 
\hline 
Zeng et al. \cite{Zeng1993} &3.40mm  &18.05s   \\ 
Beer and Best \cite{Beer1977}&3.35mm  &16.40s \\  
\hline 
\end{tabular}
\label{table:nonlin} 
\end{table}

\begin{table}[ht]
\caption{Oscillating pressure case } 
\centering 
\begin{tabular}{c c c } 
\hline\hline 
Force model & Departure diameter & Departure time \\ [0.5ex] 
\hline 
Zeng et al. \cite{Zeng1993} &3.20mm  &8.08s   \\ 
Beer and Best \cite{Beer1977}&3.10mm  &6.50s \\  
\hline 
\end{tabular}
\label{table:nonlin} 
\end{table}

\begin{figure}
\centering
\includegraphics[width=\linewidth]{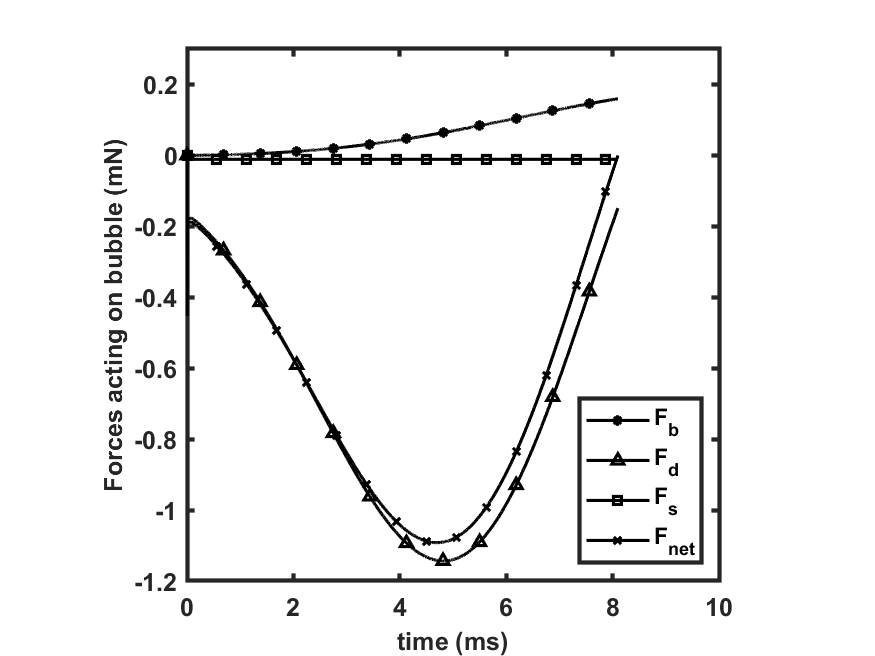}
\caption{Forces acting on the bubble during growth under an oscillating pressure field $F(t)=-5sin\omega t$, using the model of Zeng et al. \cite{Zeng1993}. }
\label{fig:forces_Klausner_sin0}
\end{figure}

\begin{figure}
\centering
\includegraphics[width=\linewidth]{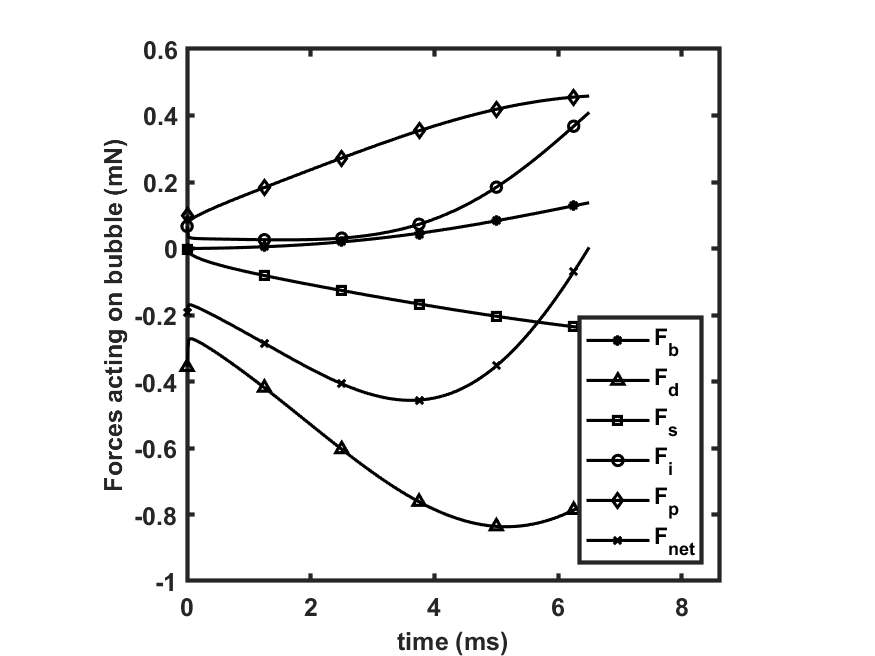}
\caption{Forces acting on the bubble during growth under an oscillating pressure field $F(t)=-5sin\omega t$, using the model of Beer and Best \cite{Beer1977}}
\label{fig:forces_Beer_sin}
\end{figure}

\begin{figure}
\centering
\includegraphics[width=\linewidth]{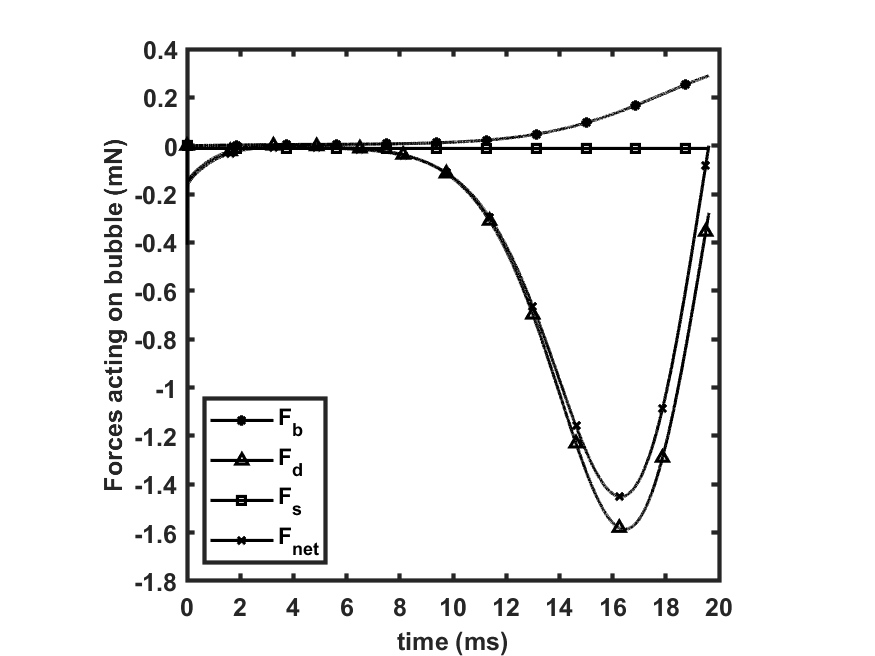}
\caption{Forces acting on the bubble during growth under an oscillating pressure field $F(t)= +5sin\omega t$, using the model of Zeng et al. \cite{Zeng1993}. }
\label{fig:forces_Klausner_sin180}
\end{figure}

\begin{figure}
\centering
\includegraphics[width=60mm,scale=0.58]{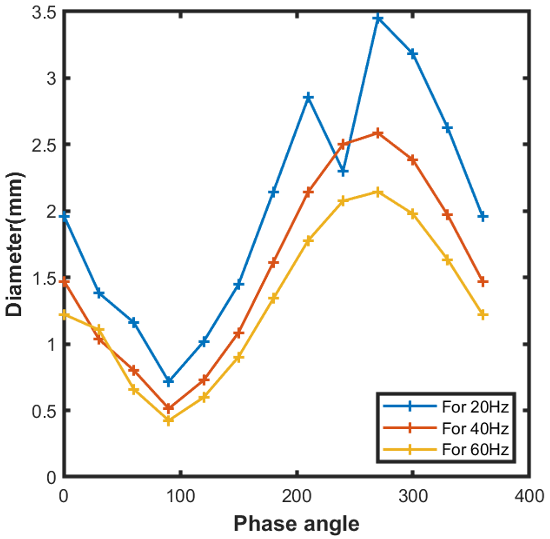}
\caption{Variation of bubble departure diameter with phase angle. Conditions are water at 1atm pressure and 10K superheat. }
\label{fig:diameter_vs_phase}
\end{figure}

\begin{figure}
\centering
\includegraphics[width=62mm,scale=0.6]{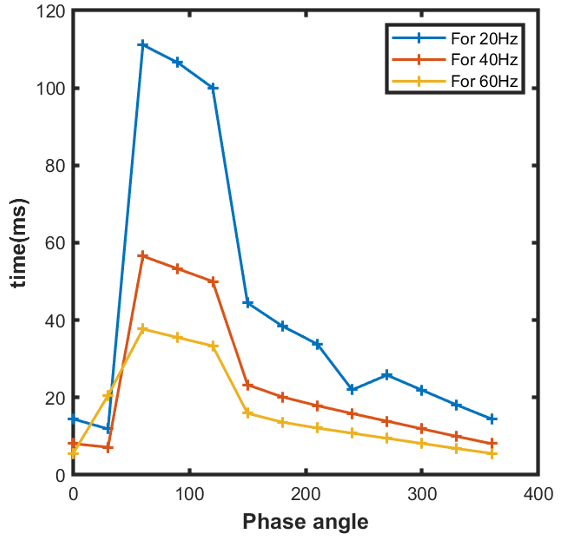}
\caption{Variation of bubble growth time as a function of initial phase angle. Conditions are water at 1atm pressure and 10K superheat.}
\label{fig:growthtime_vs_phase}
\end{figure}

The effects of initial phase angle on the bubble departure diameter and growth time after nucleation are illustrated in Figs. \ref{fig:diameter_vs_phase} and \ref{fig:growthtime_vs_phase}. It is apparent that changes in the initial phase can have significant effects on these quantities. The departure diameter appears to vary with the initial pressure phase angle in an oscillatory manner, suggesting that the time scale of the pressure fluctuation may be slow relative to the bubble growth time scale for the chosen frequencies of 40-70 Hz. The departure time, on the other hand, is a strongly nonlinear function of the initial phase angle, rising and falling steeply over a narrow range of initial phase angles. These behaviors are related to the relative dominance of the first term in the integral expression for R(t) in Eq(\ref{eq:main}) vs the second term, whose contribution to early-stage growth depends on the initial phase angle.

\begin{figure}
\centering
\includegraphics[width=\linewidth]{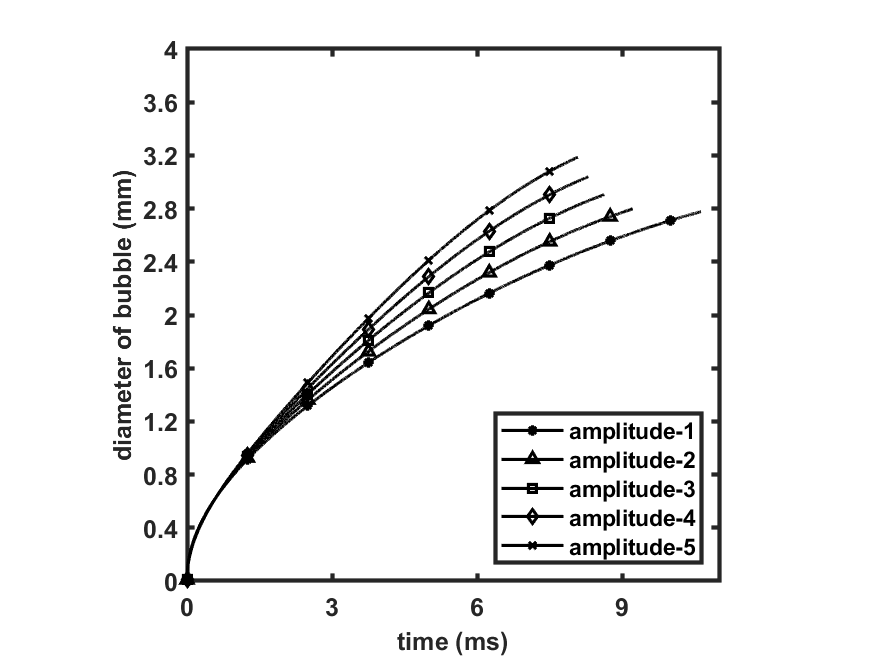}
\caption{Effect of changes in saturation temperature around the bubble due to advective effects on the thermally-controlled growth phase of the bubble.}
\label{fig:growth_ampitude}
\end{figure}

Fig. \ref{fig:growth_ampitude} shows bubble growth curves for different levels of the amplitude of the oscillation. In addition to the temperature fluctuation amplitude resulting from pressure fluctuations, the amplitude variation also accounts for the advective velocities from nearby bubbles, which may also bring cooler or hotter fluid into the superheated macrolayer surrounding the curved surface of the bubble. As expected, an increase in amplitude causes an decrease in bubble growth time, due to an increase in bubble growth rate. 

\begin{figure}
\centering
\includegraphics[width=\linewidth]{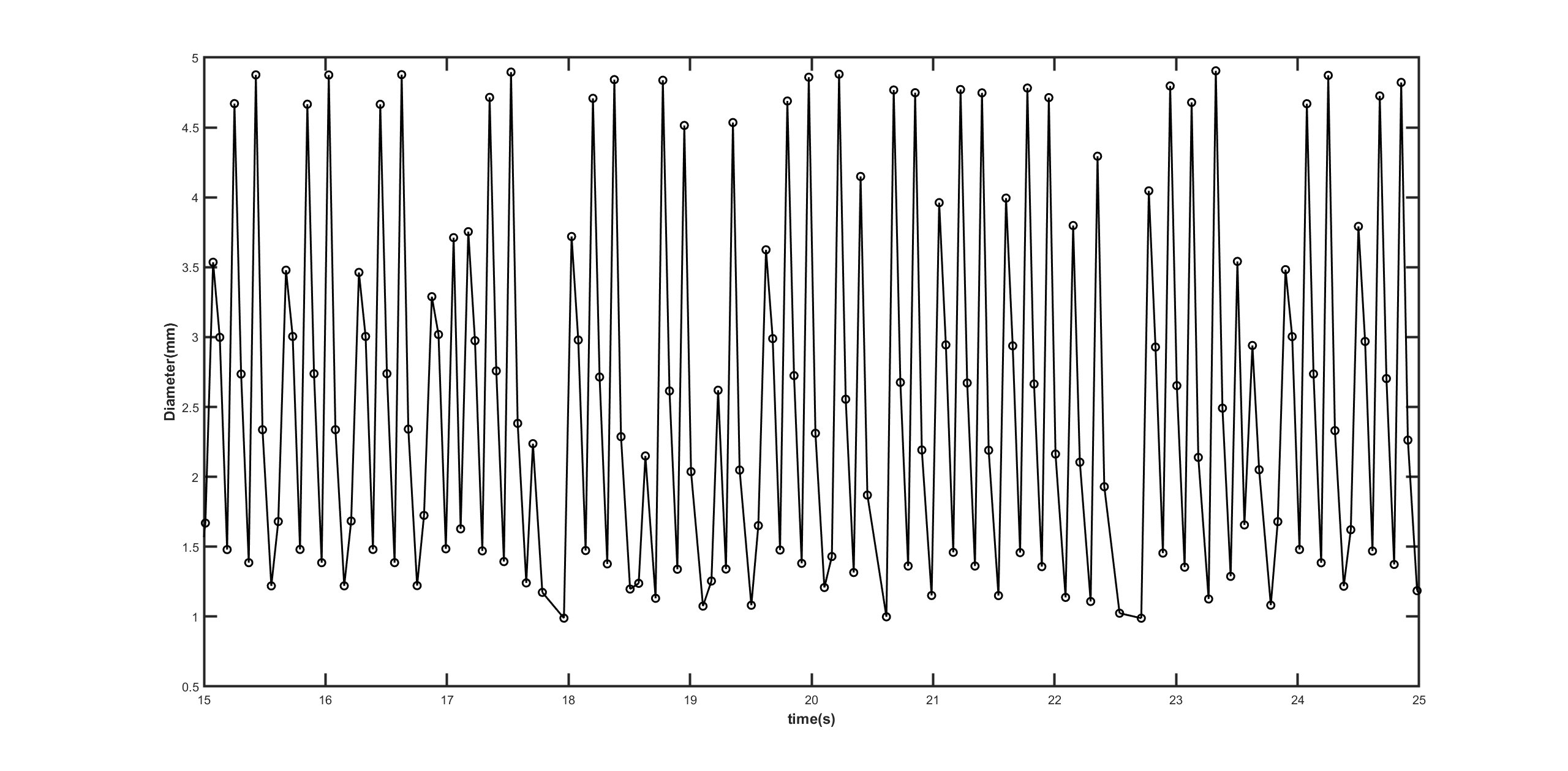}
\caption{A sequence of bubble departure events occurring over one second, when the first bubble starts growing at a pressure phase angle of $0^o$. }
\label{fig:timeseries}
\end{figure}

Putting the above results together, one can simulate a sequence of bubble departure events under an oscillating pressure field. The calculation is started at 0, with $F(t)=-Asin\omega t$, and simultaneously the growth of the first bubble is initiated. When conditions for departure are satisfied under the model of \cite{Zeng1993}, the bubble departs, a waiting time $t_w$ is given and the next bubble initiates, while the pressure clock runs continuously.  The waiting time is calculated using the expression of Han and Griffith \cite{Han1965a}
\begin{equation}\label{eq:waiting time}
    t_w = \frac{9}{4\pi \alpha} \left[  \frac{(T_w-T_\infty)R_c}{T_w - T_{sat}(1+ \frac{2\sigma}{R_c\rho _vh_{fg}}) }\right]^2
\end{equation}
The variation of waiting time causes the bubbles to lose periodicity in conjunction with a fluctuating field. This effect is stronger at higher superheat values. The initial phase angle has already been shown above to cause strong variations in the bubble growth time, which induces a change in the initial phase angle for succeeding bubbles. As a result, the sequence of departure events in Fig. \ref{fig:timeseries} shows that the bubble departure times are aperiodic, and the bubble departure diameters also become nonuniform, since each bubble now grows under a unique set of pressure conditions that are different from immediately preceding or succeeding bubbles. The result is that, instead of a fixed departure diameter time or frequency, one arrives at a probability distribution for diameter and time between departures. While a pattern in departure events may be detected by the human eye in Fig. \ref{fig:timeseries}, it was verified by performing a Fast Fourier Transform on the departure times and diameters that there was no frequency peak. 

\begin{figure}
\begin{subfigure}{\linewidth}
\includegraphics[width=0.8\linewidth]{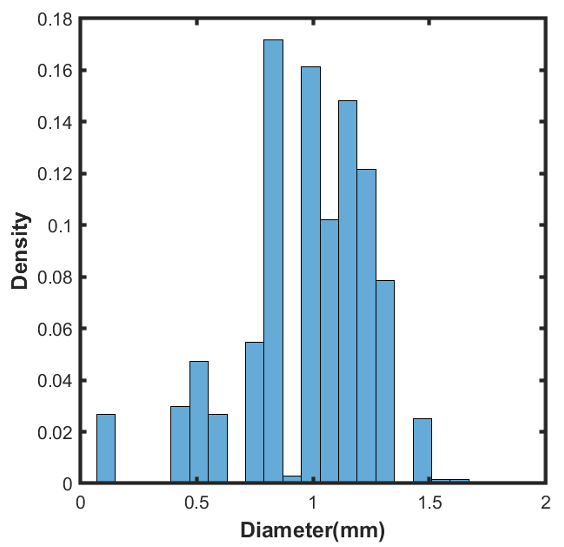}    
\caption{}
\end{subfigure}
\begin{subfigure}{\linewidth}
\includegraphics[width=0.8\linewidth]{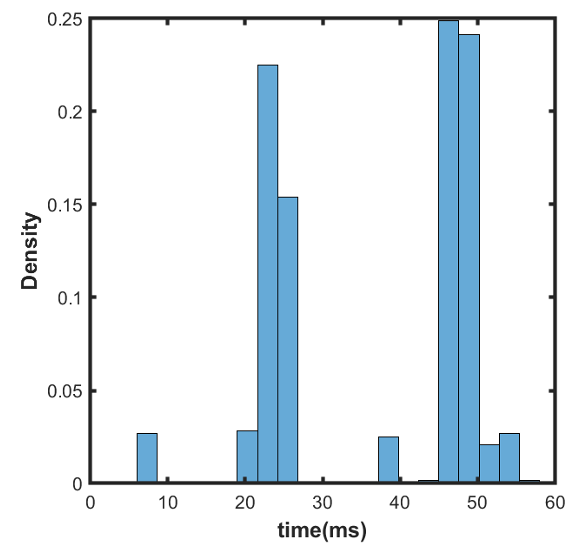}
\caption{}
\end{subfigure}
\caption{A histogram of bubble departure diameters for departure events occurring over a period of 25s. The working fluid is water at 1 atm pressure with a pressure fluctuation frequency f=40 Hz.}
\label{fig:hist_2K}
\end{figure}

\begin{figure}
\begin{subfigure}{\linewidth}
\includegraphics[width=0.98\linewidth]{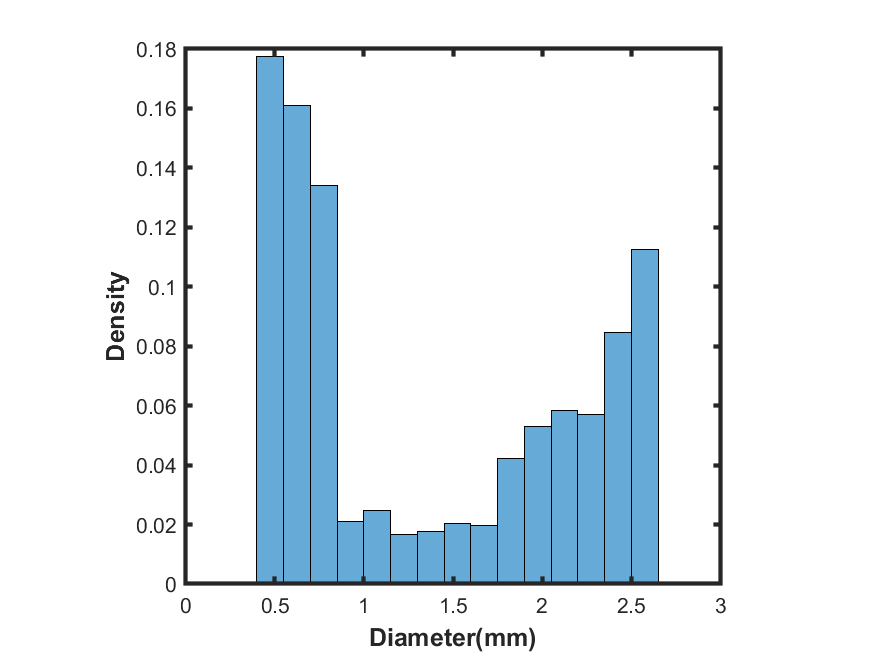}    
\caption{}
\end{subfigure}
\begin{subfigure}{\linewidth}
\includegraphics[width=0.98\linewidth]{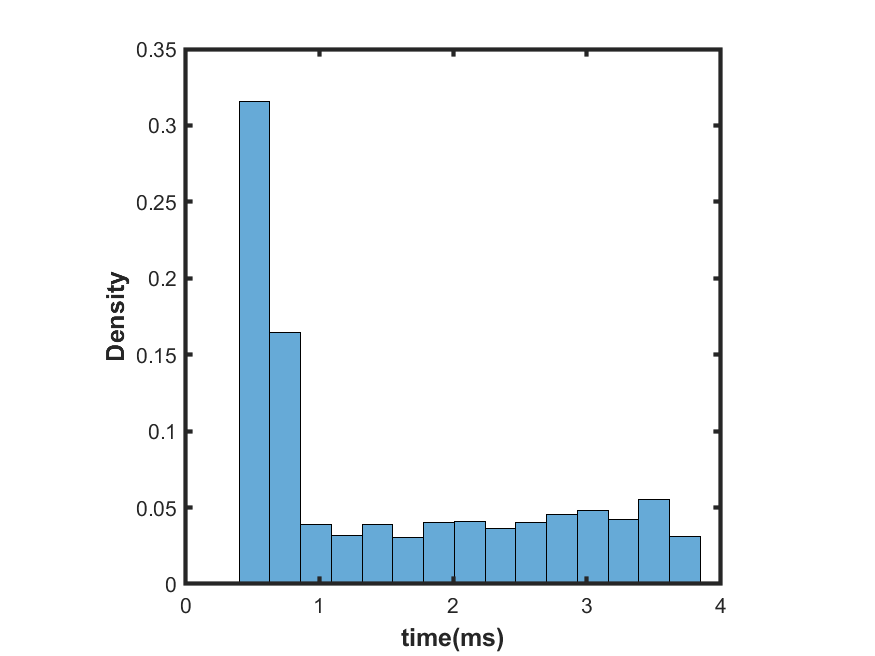}
\caption{}
\end{subfigure}
\caption{A histogram of bubble departure diameters for departure events. The working fluid is water at 1 atm pressure with a pressure fluctuation frequency f=230 Hz and 18K superheat.}
\label{fig:hist_10K}
\end{figure}

For a given fluid, this distribution is a function of wall superheat, the assumed frequency of departure of nearby bubbles, and the waiting time. Figures \ref{fig:hist_2K} and \ref{fig:hist_10K} show the probability density function for departure diameter and growth times for superheats of 10K, 18K and frequencies of 40Hz and 230 Hz respectively. At a superheat of 10K, it is observed in Fig. \ref{fig:hist_2K}(a) and (b) that the departure diameter and growth time both display a distribution. The departure diameter has values that are significantly smaller than its most probable value. The growth time behavior shows a bimodal distribution. In fact, on closer examination, it appears that the two peaks in departure time correspond to a doubling of time period of bubble departure. 

At higher superheats of 18K and higher frequencies of 230Hz, Fig. \ref{fig:hist_10K} shows that the trends in departure diameter remain similar. The bubble growth time becomes more broadly distributed in value, with the most probable value moving towards smaller values with a loss of the period-doubling behavior. It is interesting to compare these trends with the observations of the computations of Sadasivan et al. \cite{Sadasivan1995}, who note that at low applied heat fluxes, bubble ebullition cycles exhibit a period-doubling behavior, which is lost at higher heat fluxes, getting replaced by more chaotic dynamics.

These results are notable because aperiodicity appears to emerge from a fairly simple reduced-order model for an isolated bubble, despite the assumption of a single driving frequency. Such a result would have been less surprising if larger simulations such as those of Sadasivan et al. \cite{Sadasivan1995a} were performed, or if all adjacent bubble departures (i,e, the forcing function) were assumed aperiodic; the output would have reflected the input. However, the mismatch between the time scale for bubble growth dynamics, as well as the addition of a waiting time cause the harmonic field to operate in a nonlinear fashion.  It remains an open question whether such behavior has similar statistical behavior to the fluctuations observed in experiments and is the focus of future work. 

Lastly, the effect of a dynamic contact angle is added to the bubble growth model. The three-phase contact line region is characterized by contact line motion, which can increase/decrease the value of contact angle depending on the advancement/recession of the contact angle, as well as high rates of evaporation, which increases the contact angle. In the present study, only the first effect is considered; inclusion of the second effect requires detailed calculations of substrate temperature distributions including conduction effects, and is reserved for a future study. Using the asymptotic analysis of Morris \cite{Morris2001, Morris2000} for a constant-temperature surface, the contact angle is expressed as a function of the capillary number Ca, given by $Ca = \frac{\mu _l\dot{R}}{\sigma}$:
\begin{equation}
    \theta_D= \theta \pm 0.1Ca^{1/4}
\end{equation}

\begin{figure}
\centering
\vspace{2in}
\caption{Effect of using a dynamic contact angle on the bubble growth model for constant pressure. }
\label{fig:contactangle}
\end{figure}
    In the initial stages of growth, the bubble expands, and the bubble base also expands, leading to a receding contact angle(-). In the late stages of bubble growth immediately before departure, the bubble experiences a significant buoyancy force which causes the bubble base to stop expanding, and move inwards towards the center, which corresponds to an advancing contact angle(+). The value of $\dot{R}$ remains positive throughout, and the question arises as to the criterion that determines the switch from receding to advancing contact angle. It is instructive to examine experimental data on bubble behavior for guidance. Considering the experiments of \cite{Bucci2021} in which force terms were calculated using high-speed visualization of the bubble/liquid interface, one can see that the buoyancy force reaches a maximum magnitude and becomes constant after that which is also a point of maximum bubble base radius (see Fig. 3* in \cite{Bucci2021}). The present calculations employ the same rule, and the change in a sign for the Ca-dependent term is made at the point where the force balance is reached and the bubble starts to move in an upward direction. Figure \ref{fig:contactangle} shows the effects of the assumptions of constants vs variable contact angle for constant pressure case around the bubble.

\section{Conclusions}
A theoretical model has been formulated that accounts for unsteady interactions between adjacent nucleation sites in pool boiling. Only hydrodynamic interactions are considered; it has been shown that thermal interactions through substrate conduction can also be influential, but these are not considered in the present study. This is a limitation of the work; however, the intent is to explore how intermittency in quantities of interest can be introduced through small modifications of existing models. It is shown that the pressure fluctuation that may result due to pressure waves from adjacent sites can cause non-negligible changes to the degree of superheat of the liquid surrounding a bubble and cause unsteady evaporative effects that alter the growth rate of the bubble from $R \propto t^{1/2}$ behavior. The phase angle of the pressure oscillation at the start of bubble growth is shown to be critical in affecting its departure diameter and growth time. This phase angle affects two important quantities" the time-dependent superheat of the liquid around the bubble as well as the waiting time for bubble nucleation. The resulting sequence of bubble departures at a single nucleation site is shown to be intermittent, with bubble diameters spanning a wide range of values and belying the assumption of a single departure diameter. The growth time of the bubble is not constant either and decreases substantially, on the whole, from the value at constant pressure conditions. At low superheats the ebullition cycle shows period doubling behavior, with a substantial number of bubbles taking around two pressure cycles to depart. This degenerates into a broader distribution of departure times at higher superheat values. Finally, the effects of dynamic contact angle are simulated by assuming, based on observation, that the contact line starts moving inward at the point of the maximum magnitude of buoyancy force on the bubble. The dynamic contact angle   

The major finding of the study is that the inclusion of even one kind of nonlinear interaction between bubbles is sufficient to cause intermittent or aperiodic behavior depending on the quantity of interest.  The period-doubling behavior of departure events can potentially constitute a route to a chaotic regime, since nearby bubbles will now experience pressure pulses of a different frequency. This reinforces the idea that the use of a single departure diameter or departure time in calculations needs to be treated with caution, The lack of a periodic behavior and the evolution of the system towards chaos also suggests that greater attention needs to be paid to the statistics of the boiling system. From this perspective, the recent use of statistical or deep learning models becomes more appropriate. The model has several limitations in that it does not include many sources of interaction that have been addressed by other investigators; however, these can potentially be included, though at the risk of losing simplicity. One possible extension that the authors are currently working on is the inclusion of substrate conduction effects and the resulting non-uniform temperature field, which can potentially be used to alter the local evaporative flux as well as the dynamic contact angle expressions used in the present study.

\bibliographystyle{asmems4}
\bibliography{Refs}

\end{document}